\documentclass[]{IEEEtran}		
\usepackage{cite}
\usepackage{amsmath}
\usepackage{amssymb}
\usepackage{physics}
\usepackage{graphicx}
\usepackage{epstopdf}
\usepackage{url}
\usepackage{mathtools}  	
\usepackage{caption}
\usepackage{amsfonts}
\usepackage{subcaption}
\usepackage{enumerate}
\begin{document}
\title{Performance Analysis of MRC under Spatially Correlated Interference using Mixture-based Method}
\author{Arindam Ghosh\IEEEmembership{} and Harpreet S. Dhillon\IEEEmembership{} \vspace{-1cm}
\thanks{A. Ghosh is with Centre for Development of Telematics, Bengaluru, India. Email: arindam.gm@gmail.com}
\thanks{H. S. Dhillon is with the Bradley Dept. of Electical and Computer Engineering, Virginia Tech, USA. Email: hdhillon@vt.edu}
}
\maketitle

\begin{abstract}
Effect of interference correlation on wireless systems is often studied by modeling the locations of interferers as a Poisson Point Process (PPP). However, in many cases, the complicated nature of this correlation limits the analytical tractability of the PPP-based approach. For example, for an interference-aware $N$-antenna Maximum-Ratio Combining (MRC) receiver, the analytical expression for outage probability is available only for $N=2$. For $N>2$, the exact analysis using standard PPP-based approach becomes intractable because of which one has to either resort to bounds or to simulations. In this letter, we overcome this issue and derive the MRC outage probability for an arbitrary $N$ by employing the mixture-based method of modeling the correlation in the interference powers. This method offers a much simpler analytical structure to the correlated interference, thereby lending analytical tractability to such analyses. The mixture parameter ($q$) is tuned based on the matching of joint Signal-to-Interference Ratio (SIR) statistics, which results in a very accurate mixture-based result. The tightness of these approximations is verified using Monte Carlo simulations.
\end{abstract}

\begin{IEEEkeywords}
Maximum ratio combining, Poisson point process, interference correlation, mixture of random variables.
\end{IEEEkeywords}

\vspace{-0.3cm}
\section{Introduction}
Owing to its optimality under white noise \cite{brennan:2003}, MRC has become almost ubiquitous in multi-antenna systems used in the current wireless ecosystem. However, like other diversity-combining techniques, it suffers from performance losses in the presence of spatially correlated interference \cite{ganti:mhaenggi:comml:2009}. Therefore, it is important to characterize its exact performance under correlated interference, which is the main topic of this letter. 

Initial attempts towards characterizing the post-combining SIR, or equivalently the MRC outage/success probability, under spatially correlated interference include \cite{tanbourgi:2014} \cite{tanbourgi:dhillon:etel:trancom:2014}, which model the interference field as PPP and employ tools from stochastic geometry for the analysis. However, due to the structure of the problem, the exact analysis was limited to only the case of $N=2$. For the general case of $N>2$, only simple bounds were provided.    

Therefore, in this letter, our objective is to characterize, or tightly approximate, the Cumulative Distribution Function (CDF) of the interference-affected post-combining SIR for arbitrary $N$. To achieve this, we employ the mixture-based method \cite{arindam:comml:2017} of modeling the correlation in the interference powers experienced at multiple antennas. This method, at its core, employs a correlation framework constructed using \textit{mixture of random variables}, which not only offers a much more amenable analytical structure but also mimics the PPP characteristics quite well. The derived analytical expressions for MRC outage probability are shown to tightly approximate the exact PPP-based results obtained using Monte Carlo simulations. 

As a part of this contribution, we also provide a method for improving the accuracy of the mixture-based approximation method of \cite{arindam:comml:2017}. Originally, in \cite{arindam:comml:2017}, the mixture weights ($q$'s) were chosen so as to match the interference correlation of the mixture-based and PPP-based models. Different from \cite{arindam:comml:2017}, in this letter, we show that tuning $q$ instead to match the joint SIR statistics, i.e. joint complementary CDF (CCDF) of SIR, of the two models results in much tighter approximations. This tuning requires solving an $N^{\text{th}}$ degree polynomial equation in $q$, which can be easily accomplished using commonly available numerical packages.

\vspace{-0.25cm}
\section{System Model}
We consider a Poisson dipole network \cite{Baccelli:Blaszczyszyn:Muhlethaler2006} wherein the transmitters (single antenna) are distributed as per a homogeneous PPP $\Phi \equiv \{x\} \subset \mathbb{R}^{2}$ of intensity $\lambda$. Each transmitter has, at a distance $d$ in an arbitrary direction, an $N$-antenna receiver capable of MRC processing. By Slivnyak's theorem \cite{mhaenggi:ganti:2009}, we then add a new reference transmitter into the network with its receiver placed at the origin $o$, which can be referred to as the ``typical" pair. In this letter, we focus on this pair to study the perfomance of the considered MRC.

We denote the location of the typical receiver's antennas by $\{z_i\}_{i=1}^{N}\in \mathbb{R}^{2}$. The separation between these antennas is assumed to be negligible (of the order of wavelength) in comparison to other distances in the network (such as $d$). Therefore, for the purpose of path-loss calculations, we will assume that all antennas are co-located with the receiver itself, i.e. $\{z_i \equiv o\}$. That said, it should be noted that the small separation between the antennas may still allow them to experience significantly different fading characteristics.  

For the typical receiver, the nodes at $\Phi$ (not including the serving transmitter) act as interferers. These interferers are assumed to follow ALOHA protocol whereby each interferer transmits independently on the same time-frequency resource block as the typical link with probability $p$. Therefore, the set of active interferers $\hat{\Phi}$ that interfere with the typical receiver in any given time slot is simply a thinned PPP of intensity $\lambda p$ (parent PPP $\Phi$ thinned with probability $p$). For channel fluctuations, we assume i.i.d. Rayleigh fading across all the links. For path-loss, between $x$ and $i$-th antenna ($z_i$), we consider the standard function $\ell(x,z_i)=\frac{1}{\epsilon+\|x-z_i\|^{\alpha}}$, where $\alpha>2$ is the path-loss exponent and $\epsilon \rightarrow 0$. For the case of $z_i \equiv o$, we simply have $\ell(x,z_i)=\frac{1}{\epsilon+\|x\|^{\alpha}}$. Lastly, we assume that all nodes in the system transmit with $P$ units of power.     

Then, in this setting, the SIR at the $i$-th antenna due to a transmission from the reference transmitter is given by
\vspace{-0.1cm}
\begin{equation}
\text{SIR}_i = \frac{Ph_{i}d^{-\alpha}}{\sum_{x\in\hat{\Phi}}^{}Ph_{xz_i}\|x\|^{-\alpha}}=\frac{h_{i}d^{-\alpha}}{I(z_i)},
\label{SIR}
\end{equation}
where, $I(z_i)$ is the interference power normalized by $P$, and $h_{xz_i}$, $h_i$ $\sim \exp(1)$ are the exponentially distributed fading gains of the link between $x$ and $z_i$ and between the reference transmitter and $i$-th antenna, respectively. We focus on interference-limited scenario in which thermal noise is negligible compared to interference. Since all antennas of the receiver of interest are collocated from the path-loss perspective and all links experience i.i.d. Rayleigh fading, it is easy to argue that the interference power experienced across different antennas is identically distributed, i.e., $\{I(z_i)\}$ is a sequence of identically distributed random variables. Further, these interference powers exhibit spatial correlation because of the common interference field. The spatial correlation coefficient $\zeta_{ij} = \text{Corr}[I(z_i), I(z_j)]$, $\forall i\neq j$, can be readily quantified using only the spatial aspect of \cite[(11)]{ganti:mhaenggi:comml:2009}, i.e.
\begin{equation}
\begin{split}
\zeta_{ij} &=\lim\limits_{\epsilon \rightarrow 0}\frac{\int_{\mathbb{R}^2}^{}\ell(x,o)^2\dd x}{\mathbb{E}[h_{xz_i}^2]\int_{\mathbb{R}^2}^{}\ell(x,o)^2\dd x}\\
& = \lim\limits_{\epsilon \rightarrow 0}\frac{\int_{\mathbb{R}^2}^{}\frac{1}{(\epsilon+\|x\|^{\alpha})^{2}}\dd x}{\mathbb{E}[h_{xz_i}^2]\int_{\mathbb{R}^2}^{}\frac{1}{(\epsilon+\|x\|^{\alpha})^{2}}\dd x}=\lim\limits_{\epsilon \rightarrow 0}\frac{1}{\mathbb{E}[h_{xz_i}^2]} = \frac{1}{2},
\end{split}
\label{zeta_same_point}
\end{equation}
where, the last step follows from $\mathbb{E}[h^n] = n!$ for $h\sim \exp(1)$.

For MRC, we consider the interference-aware case, as also assumed in \cite{tanbourgi:2014} \cite{tanbourgi:dhillon:etel:trancom:2014}, where the receiver has perfect knowledge of the instantaneous interference powers and the transmitter-to-receiver link fading gains for every antenna. The combiner treats the interference as white noise and the MRC weights are taken to be proportional to the ratio of fading amplitude and interference power \cite{brennan:2003}. The post-MRC SIR, denoted by $\text{SIR}_\text{MRC}$, is therefore given by 
\begin{equation}
\text{SIR}_\text{MRC} =\frac{h_{1} d^{-\alpha}}{I(z_1)} + \ldots + \frac{h_{N} d^{-\alpha}}{I(z_N)}.
\label{sir_mrc_expansion}
\end{equation}
For the case of $N=2$, this SIR has been fully characterized in \cite{tanbourgi:2014}. However, for the general case, the standard approaches are not analytically tractable, which makes the analysis extremely difficult. In the next section, we present an alternate approach, wherein we employ the mixture-based method of interference modeling \cite{arindam:comml:2017}, to obtain the CDF of $\text{SIR}_\text{MRC}$ for arbitrary $N$.

\vspace{-0.15cm}     
\section{Outage Probability of MRC}
The mixture-based model in \cite{arindam:comml:2017} can be used to construct a set of arbitrarily (non-negatively) correlated interference powers $\{I(z_i)\}_{i=1}^N$. For the multi-antenna case of this letter, where $I(z_i)$ and $I(z_j)$ are equally correlated $\forall i\neq j$, the mixture-based framework of \cite{arindam:comml:2017} can be reduced to a more simpler form as presented in the following.

\vspace{-0.4cm}  
\subsection{Mixture-based Construction of Interference Powers}
The main idea is to represent the PPP-based interference random variables $\{I(z_i\equiv o)\}_{i=1}^N$ with their mixture-based equivalents having the same PPP-based distributional and correlational properties. To achieve this, we first generate a sequence of i.i.d. random variables $\{J_n\}_{n=0}^{N}$ that are distributed identically to $I(o)$ of (1). For that, we imagine an auxiliary setup wherein we take a set of independent homogeneous PPPs of interferers $\{{\Psi}_n\}_{n=0}^N$, of intensity $\lambda p$, on $\mathbb{R}^2$ and model $\{J_n\}$ to be the respective interference powers observed at the origin $o$ due to these PPPs, i.e.
\begin{equation}
J_{n} = \sum_{x\in{\Psi}_{n}}^{}h_{xo}\ell(x,o) \qquad \text{for}\ n\in\{0,\ldots,N\}.
\label{mixture_rv_J}
\end{equation}
Note that $\{{\Psi}_n\}$, which are used to obtain $\{J_n\}$, belong only to the auxiliary system and in no way interfere with the actual PPP $\hat{\Phi}$ of the interferers in the original system of Section II.

Next, using these $J_n$'s, we model $\{{I}(z_i)\}$ as the mixtures  
\begin{equation}
\begin{split}
I(z_i) = J_{A_i}, \qquad \text{for}\ i\in\{1,\ldots,N\},
\label{I_mixture}
\end{split}
\end{equation}
where, $\{A_i\}_{i=1}^{N}$ are independent binary random variables whose probability mass functions are given by
\begin{equation}
p_{A_i}(a_i)=
\begin{cases}
	q, & \text{if}\ a_i=0 \\
	1-q, & \text{if}\ a_i=i\\
	0, & \text{otherwise}.
\end{cases}
\end{equation}
From the distribution preservation property of mixtures [5], we have the mixture-based $I(z_i)$ distributed identically to $J_0$ and $J_i$ and therefore to PPP-based $I(z_i)$ $\forall i$. Further, it can be easily verified that the mixture-based $\text{Corr}[{I}(z_i),{I}(z_j)]$ $=$ $q^2$, for all $i\neq j$. At this point, it appears natural to select $q^2 = 0.5$ so as to match the values of the correlation coefficient with (2), but, as discussed in the sequel, this does not necessarily offer the most accurate approximation. Next, we use this mixture-based framework to derive the MRC outage probability for an arbitrary $N$.
\vspace{-0.3cm}  
\subsection{CDF of $\text{SIR}_\text{MRC}$ for arbitrary $N$}
The main result of this paper is stated in the next Theorem.

\noindent\textbf{Theorem 1} \textit{The outage probability of MRC, defined as $\text{P}_{\text{Out}} \stackrel{\Delta}{=}\mathbb{P}(\text{SIR}_\text{MRC} < T)$, derived using the mixture-based method, is given by \eqref{eq_no_big_Pout}, where $T$ is the given threshold, $C=\frac{2\pi^2\lambda p}{\alpha \sin(2\pi/\alpha)}$, $f_{V_k}(v_k) = \frac{2C}{\alpha}v_k^{\frac{2}{\alpha}-1}e^{-C v_k^{2/\alpha}}$, and $B_{m,j}(\cdot)$ is the Bell polynomial \cite{jhonson:bell}.}
 
\textit{Proof:} From \eqref{sir_mrc_expansion}, we have
\vspace{-0.4cm}
\begin{figure*}[!t]
	\normalsize
	\begin{equation*}
	\begin{split}
	\text{P}_{\text{Out}} &=\sum_{n=0}^{N} {N \choose n} q^n (1-q)^{N-n}\hspace{-0.15cm}\int_{0}^{Td^{\alpha}}\hspace{-0.3cm}\int_{0}^{Td^{\alpha}\hspace{-0.1cm}-v_1}\hspace{-0.7cm}\cdots \int_{0}^{Td^{\alpha}\hspace{-0.1cm}-v_1\hspace{-0.05cm}-\ldots -v_{N-n-1}}\hspace{-0.1cm}\Biggl\{\hspace{-0.05cm}1\hspace{-0.05cm}-\hspace{-0.17cm}\sum_{m=0}^{n-1}\hspace{-0.075cm}(-1)^m\hspace{-0.05cm}\frac{\left(\hspace{-0.05cm}Td^{\alpha}\hspace{-0.15cm}-\hspace{-0.1cm}\sum_{k=1}^{N-n} \hspace{-0.1cm}v_k\hspace{-0.05cm}\right)^{\hspace{-0.1cm}m}}{m!}\hspace{-0.15cm}\sum_{j=1}^{m}\hspace{-0.05cm}(\hspace{-0.05cm}-1\hspace{-0.05cm})^je^{-C \left(\hspace{-0.05cm}Td^{\alpha}\hspace{-0.1cm}-\hspace{-0.05cm}\sum_{k=1}^{N-n} \hspace{-0.075cm}v_k\hspace{-0.05cm}\right)^{\frac{2}{\alpha}}}\hspace{-0.1cm}.\\
	&\hspace{1.2cm}B_{m,j}\biggl\{\hspace{-0.05cm}\frac{2C\hspace{-0.05cm}\left(\hspace{-0.1cm}Td^{\alpha}\hspace{-0.15cm}-\hspace{-0.175cm}\sum_{k=1}^{N-n} \hspace{-0.1cm}v_k\hspace{-0.1cm}\right)^{\hspace{-0.1cm}\frac{2}{\alpha}\hspace{-0.025cm}-\hspace{-0.025cm}1\hspace{-0.5cm}}}{\alpha}, \ldots, \hspace{-0.075cm}\frac{\hspace{-0.05cm}2\hspace{-0.05cm}\left(\hspace{-0.05cm}2\hspace{-0.075cm}-\hspace{-0.075cm}\alpha\hspace{-0.05cm}\right)\hspace{-0.05cm}\cdots\hspace{-0.05cm} \left(\hspace{-0.05cm}2\hspace{-0.075cm}-\hspace{-0.075cm}\alpha(\hspace{-0.05cm}m\hspace{-0.075cm}+\hspace{-0.075cm}j\hspace{-0.05cm})\hspace{-0.05cm}\right)\hspace{-0.05cm}C\hspace{-0.05cm}\left(\hspace{-0.1cm}Td^{\alpha}\hspace{-0.15cm}-\hspace{-0.175cm}\sum_{k=1}^{N-n}\hspace{-0.1cm}v_k\hspace{-0.1cm}\right)^{\hspace{-0.1cm}\frac{2}{\alpha}\hspace{-0.05cm}-\hspace{-0.05cm}1\hspace{-0.05cm}-\hspace{-0.05cm}m\hspace{-0.05cm}+\hspace{-0.05cm}j\hspace{-0.8cm}}}{\alpha^{m-j+1}}\biggr\}\hspace{0.4cm}\Biggr\}\Biggl\{\prod_{k=1}^{N-n}\hspace{-0.075cm}f_{V_k}\hspace{-0.05cm}(v_k)\hspace{-0.05cm}\Biggr\}\dd v_{N-n}\ldots \dd v_1
	\end{split}
	\tag{15} \label{eq_no_big_Pout}
	\end{equation*}
	\hrulefill
	\vspace*{-0.4cm}
\end{figure*}

\begin{equation*}
\begin{split}
\text{P}_{\text{Out}} &= \mathbb{P}\left(\frac{h_{1} d^{-\alpha}}{I(z_1)} + \ldots + \frac{h_{N} d^{-\alpha}}{I(z_N)} < T\right)\\
&\stackrel{(b)}{=}\mathbb{P}\left(\frac{h_1}{J_{A_1}} + \ldots + \frac{h_N}{J_{A_N}} < Td^{\alpha}\right)\\
&\stackrel{(c)}{=} \sum_{a_1\in \{0,1\}}^{} \sum_{a_2\in \{0,2\}}^{} \ldots \sum_{a_N \in \{0,N\}}^{} \left(\prod_{i=1}^{N}p_{A_i}(a_i)\right).\\
&\ \ \ \mathbb{P}\left(\frac{h_1}{J_{A_1}} + \ldots + \frac{h_N}{J_{A_N}} < Td^{\alpha}\Bigr\rvert A_1=a_1,\ldots,A_N=a_N\right)
\end{split}
\end{equation*}
where, (b) follows from the mixture-based representations of $\{{I}(z_i)\}$ in \eqref{I_mixture}, and $(c)$ from the law of total probability. Next, by collecting the common terms, the above can be compactly written as: $\text{P}_{\text{Out}} =$
\begin{equation}
\begin{split}
\hspace{-0.1cm}\sum_{n=0}^{N} \hspace{-0.1cm}{N \choose n} q^n (1\hspace{-0.075cm}-\hspace{-0.075cm}q)^{N-n}\underbrace{\mathbb{P}\left(\frac{\sum_{i=1}^{n}h_i}{J_0} + \sum_{k=1}^{N-n}\frac{h_k}{J_k} < Td^{\alpha}\right)}_{=W_n}
\end{split}
\label{P_out}
\end{equation}
Here, we define $U_n=\frac{\sum_{i=1}^{n}h_i}{J_0}$ and $V_k=\frac{h_k}{J_k}$ and find their distributions. In this regard, we observe that $U_n d^{-\alpha}$ is equivalent to the random variable $\text{SIR}_\text{MRC}$ when all the $N$ antennas see the same interference power. This has been studied as the full-correlation case in \cite{tanbourgi:2014}, and therefore, from \cite[Lemma 1, Proposition 1]{tanbourgi:2014}, we have the CDF of $U_n$ given by
\vspace{-0.4cm}
\begin{equation}
F_{U_n}(u) = 1-\sum_{m=0}^{n-1}(-1)^m\frac{u^m}{m!}\frac{\partial^m }{\partial u^m}\exp\left(-C u^{2/\alpha}\right),
\label{Un_cdf}
\end{equation}
where $C=\frac{2\pi^2\lambda p}{\alpha \sin(2\pi/\alpha)}$. Next, note that $V_k$ is distributed identically to $U_1$; hence, from \eqref{Un_cdf}, we have the CDF and PDF of $V_k$, respectively, given by \vspace{-0.2cm}
\begin{align}
F_{V_k}(v_k)& = 1-\exp\left(-C v_k^{2/\alpha}\right),\ \  \text{and}\\
f_{V_k}(v_k)& = \frac{2C}{\alpha}v_k^{\frac{2}{\alpha}-1}\exp\left(-C v_k^{2/\alpha}\right).
\label{Vk_cdf_pdf}
\end{align}
\vspace{-0.3cm}
Using these distributions, $W_n$ can now be derived as
\begin{equation}
\begin{split}
W_n &= \mathbb{P}\left(U_n +  \sum_{k=1}^{N-n} V_k < Td^{\alpha}\right)\\
&=\mathbb{E}_{\{V_k\}}\left[\mathbb{P}\left(U< Td^{\alpha}-\sum_{k=1}^{N-n} V_k\right)\right]\\
&=\mathbb{E}_{\{V_k\}}\left[\left.1-\sum_{m=0}^{n-1}(-1)^m\frac{s^m}{m!}\frac{\partial^m }{\partial s^m}\exp\left(-C s^{2/\alpha}\right)\right\rvert_{s}\right],
\end{split}
\end{equation}
where $\rvert_{s}$ means the function inside the expectation is evaluated at $s=Td^{\alpha}-\sum_{k=1}^{N-n} V_k$. Next, using Faa di Bruno's formula \cite{bruno:1857} and Bell polynomial $B_{m,j}(\cdot)$ \cite{jhonson:bell}, for the $m$-th derivative, we get
\vspace{-0.4cm}
\begin{equation}
\begin{split}
W_n&=\mathbb{E}_{\{V_k\}}\biggl[1-\sum_{m=0}^{n-1}(-1)^m\frac{s^m}{m!}\sum_{j=1}^{m}(-1)^je^{-C s^{\frac{2}{\alpha}}}.\\
&\qquad\qquad B_{m,j}\left\{\frac{\partial}{\partial s}C s^{\frac{2}{\alpha}}, \ldots, \frac{\partial^{m-j+1}}{\partial s^{m-j+1}}C s^{\frac{2}{\alpha}} \right\}\biggr\rvert_{s}\biggr],
\end{split}
\end{equation} 
Finally, averaging over the i.i.d. $\{V_k\}$ and substituting into \eqref{P_out}, we get the result.

\vspace{-0.3cm}
\subsection{Tuning of the Mixture Parameter ``$q$"}\label{mapping_ssec}

We conclude the mixture-based model of this letter by choosing a value for the parameter $q$ that results in highly accurate mixture-based approximations. Here, we propose a new method of tuning (based on matching of joint statistics of the SIR) which offers a much tighter approximation than the tuning presented originally in \cite{arindam:comml:2017}. Recall that tuning in \cite{arindam:comml:2017} is based on interference correlation matching, which results in $q^2=0.5$ from \eqref{zeta_same_point}.

The MRC outage probability involves (jointly) multiple correlated SIRs; and, as the accuracy of the mixture-based approximation depends on how closely it mimics the PPP-based characteristics, we propose to match the two models directly at the level of joint SIR statistics. That is, tune $q$ such that the joint CCDFs of SIR match. 
The difference between the two CCDFs, for a threshold $T$, is given by
\begin{equation}
\begin{split}
&f(q) = \ e^{\frac{-B\Gamma\left(N+\frac{2}{\alpha}\right)}{(N-1)!\ \Gamma\left(1+\frac{2}{\alpha}\right)}}\\
&- \sum_{n=0}^{N}{N \choose n}q^n(1-q)^{N-n}\exp{-B(n^{\frac{2}{\alpha}}+N-n)},
\end{split}
\label{eq_joint_ccdf_matching}
\end{equation}
where, $B=\frac{2\pi^2\lambda p d^2 T^{\frac{2}{\alpha}}}{\alpha \sin(2\pi/\alpha)}$. The first CCDF term is PPP-based that is available from \cite[Theorem 1]{haenggi:2012}. The second term is for the mixture-based case, which is given as: $\mathbb{P}\left(\text{SIR}_1>T, \ldots,\text{SIR}_N>T\right)$ =
\begin{equation*}
\begin{split}
& \mathbb{P}\left(\cap_{i=1}^Nh_i>yd^{\alpha}I(z_i)\right)= \mathbb{E}_{\{J_{A_i}\}}\biggl[e^{-\sum_{i=1}^{N}yd^{\alpha}J_{A_i}}\biggr]\\
&\stackrel{(b)}{=} \hspace{-0.25cm}\sum_{a_1\in \{0,1\}}^{} \hspace{-0.25cm}\ldots\hspace{-0.2cm} \sum_{a_N \in \{0,N\}}^{} \hspace{-0.15cm}\left(\prod_{i=1}^{N}p_{A_i}(a_i)\right)\mathbb{E}_{\{J_{a_i}\}}\biggl[e^{-\sum_{i=1}^{N}yd^{\alpha}J_{a_i}}\biggr]\\
&\stackrel{(c)}{=}\sum_{n=0}^{N} {N \choose n} q^n (1\hspace{-0.1 cm}-\hspace{-0.1 cm}q)^{N\hspace{-0.05 cm}-\hspace{-0.05 cm}n}\mathbb{E}_{J_0}\hspace{-0.05 cm}\biggl[e^{-nyd^{\alpha}J_{0}}\biggr]\mathbb{E}_{J_1}\hspace{-0.05 cm}\biggl[e^{-yd^{\alpha}J_{1}}\biggr]^{\hspace{-0.05 cm}N-n},
\end{split}
\end{equation*}
where, (b) follows from the total probability law and (c) by collecting the common terms. Lastly, by taking the Laplace transform of interference \cite{mhaenggi:ganti:2009}, we get the expression in \eqref{eq_joint_ccdf_matching}. 

The tuned $q$ which matches the two CCDFs is simply the solution of the equation $f(q)=0$. From Abel-Ruffini's theorem \cite{jacobson:2009}, there is no general algebraic solution for polynomial equations with arbitrary coefficients for $N\geq 5$. Although this does not mean that some particular classes of higher degree polynomials may not have explicit form of algebraic solutions, such an exploration for the above equation is outside the scope of this study. Therefore, we simply use commonly available numerical packages from Mathematica or Matlab to find the appropriate value of $q\in[0,1]$ which satisfies the above equation. 

The above tuning, as shown in the next section, results in far better accuracy compared to simply setting $q^2=\zeta_{ij}=0.5$ as per \cite{arindam:comml:2017}. This is because even if the two models are matched at the interference correlation level, deviations may still appear at the SIR correlation level. To understand this, note that the SIR correlation can be expressed as
\begin{equation}
\text{Corr}[\text{SIR}_i,\text{SIR}_j] = \zeta^{\text{inv}}_{ij}\sqrt{\frac{\text{Var}[I(z_i)^{-1}]\text{Var}[I(z_j)^{-1}]}{\text{Var}[h_iI(z_i)^{-1}]\text{Var}[h_jI(z_j)^{-1}]}}
\end{equation}
where, $\zeta^{\text{inv}}_{ij}= \text{Corr}\left[
I(z_i)^{-1},I(z_j)^{-1}\right]$. From results obtained using Monte Carlo simulations, we find that for an interference correlation of $0.5$ (for $\alpha=4$ and $\lambda p=10^{-2}/\text{m}^2$), $\zeta^{\text{inv}}_{ij}\approx 0.76$, which is then brought down to the SIR correlation of $\approx 0.3$. However, for mixtures, it can be easily shown that if $\zeta_{ij}=0.5$, $\zeta^{\text{inv}}_{ij}$ is also $0.5$, which therefore will result in an SIR correlation of $< 0.3$. Clearly, this mismatch will affect the accuracy of the derived approximations, which motivated us to tune $q$ as per \eqref{eq_joint_ccdf_matching} instead.

\vspace{-0.2cm}
\section{Simulation Results}\label{simulation_sec}

\begin{figure}[t]
\centering
\includegraphics[scale=0.46]{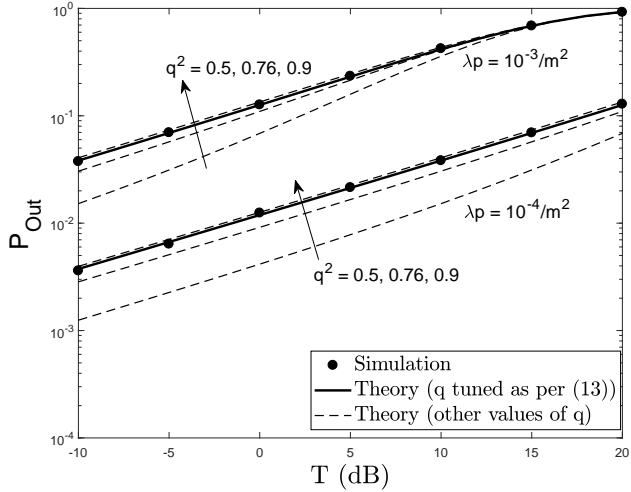}
\caption{$P_{\text{Out}}$ vs. $T$ for $N=4$, $\alpha=4$, and $d=10$m.}
\label{Pout_vs_T_N4}
\vspace{-0.5cm}
\end{figure}

\begin{figure}[t]
\centering
\includegraphics[scale=0.45]{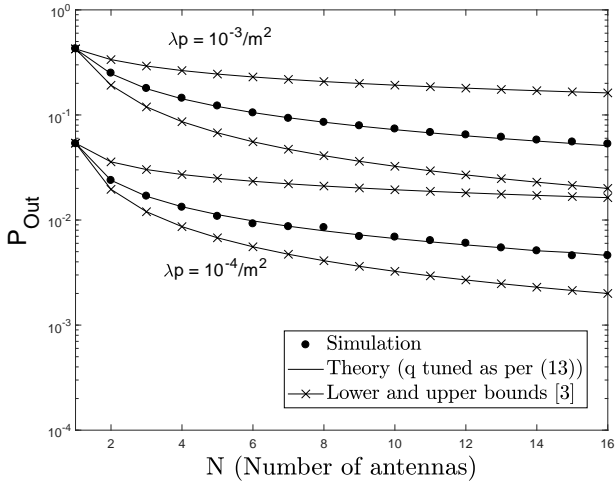}
\caption{$P_{\text{Out}}$ vs. $N$, $T=1$dB, $\alpha=4$, $d=10$m.}
\label{Pout_vs_N}
\vspace{-0.5cm}
\end{figure}

\begin{figure}[t]
\centering
\includegraphics[scale=0.47]{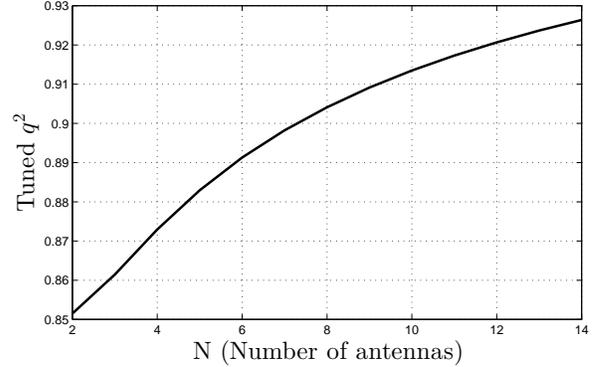}
\caption{Tuned $q^2$, $T=1$ dB, $\lambda p=10^{-4}/\text{m}^2$, $\alpha=4$, $d=10$m.}
\label{fig_tuned_q_values}
\vspace{-0.5cm}
\end{figure}

For simulations, we consider a square region $[-L,L]^2$ centered at the origin such that there are on an average 1000 interferers present, i.e. $L$ satisfies $(2L)^2\lambda p=10^3$. The receiver is then placed at the center with its reference transmitter at a distance $d=10$m. For path-loss, we take $\alpha=4$. 

Figure \ref{Pout_vs_T_N4} plots the outage probability of the $4$-antenna MRC receiver with respect to the threshold $T$. Clearly, the accuracy of the mixture-based approximations depends on the value of $q$. The tightest match with the simulation data is obtained when $q$ is tuned as per \eqref{eq_joint_ccdf_matching}. Setting $q^2=\zeta^{\text{inv}}_{ij}=0.76$ (which matches the two models at the SIR correlation level) gives better accuracy than $q^2=\zeta_{ij}=0.5$, however, it still slightly undershoots the plot. A very high value of $q^2$ ($=0.9$), on the other hand, causes overshooting.  

Similar behavior is observed in Fig. \ref{Pout_vs_N}, which plots $\text{P}_{\text{Out}}$ against the number of receiver antennas. We see that the tuned values of $q$ (plotted in Fig. \ref{fig_tuned_q_values}), provide the tightest approximations to the actual PPP-based data. For comparison purpose, we also plot the upper and lower bounds of the MRC outage probability that were proposed in \cite[Proposition 2]{tanbourgi:2014} for arbitrary $N$. Not surprisingly, the mixture-based approximations offer far accurate results than the PPP-based bounds.   

\vspace{-0.15cm}
\section{Conclusions}
In this letter, using the mixture-based method of modeling spatially correlated interference, we derived accurate expressions for the MRC outage probability for any arbitrary number of antennas, which was earlier not possible through standard PPP-based approach. In addition, different from \cite{arindam:comml:2017}, we present a new method of tuning the mixture parameter $q$ that is based on matching the joint CCDF of SIR. These tuned mixture-based approximations are shown to obtain far better accuracy than the tuning in \cite{arindam:comml:2017} and the previously known bounds in the literature. The mixture-based model in this letter, therefore, can be used to study many such related scenarios of correlated interference in multi-antenna systems.
\vspace{-0.2cm}
\bibliographystyle{IEEEtran}
\bibliography{bib_mrc_paper}

\begin{thebibliography}{10}
\providecommand{\url}[1]{#1}
\csname url@samestyle\endcsname
\providecommand{\newblock}{\relax}
\providecommand{\bibinfo}[2]{#2}
\providecommand{\BIBentrySTDinterwordspacing}{\spaceskip=0pt\relax}
\providecommand{\BIBentryALTinterwordstretchfactor}{4}
\providecommand{\BIBentryALTinterwordspacing}{\spaceskip=\fontdimen2\font plus
\BIBentryALTinterwordstretchfactor\fontdimen3\font minus
  \fontdimen4\font\relax}
\providecommand{\BIBforeignlanguage}[2]{{%
\expandafter\ifx\csname l@#1\endcsname\relax
\typeout{** WARNING: IEEEtran.bst: No hyphenation pattern has been}%
\typeout{** loaded for the language `#1'. Using the pattern for}%
\typeout{** the default language instead.}%
\else
\language=\csname l@#1\endcsname
\fi
#2}}
\providecommand{\BIBdecl}{\relax}
\BIBdecl

\bibitem{brennan:2003}
D.~G. Brennan, ``Linear diversity combining techniques,'' \emph{Proc. of the
  IEEE}, vol.~91, no.~2, pp. 331--356, Feb 2003.

\bibitem{ganti:mhaenggi:comml:2009}
R.~K. {Ganti} and M.~{Haenggi}, ``Spatial and temporal correlation of the
  interference in {ALOHA} ad hoc networks,'' \emph{{IEEE Commun. Letters}},
  vol.~13, no.~9, pp. 631 -- 633, Sept. 2009.

\bibitem{tanbourgi:2014}
R.~Tanbourgi, H.~S. Dhillon, J.~G. Andrews, and F.~K. Jondral, ``Effect of
  spatial interference correlation on the performance of maximum ratio
  combining,'' \emph{IEEE Trans. on Wireless Commun.}, vol.~13, no.~6, pp.
  3307--3316, June 2014.

\bibitem{tanbourgi:dhillon:etel:trancom:2014}
R.~{Tanbourgi}, H.~S. {Dhillon}, J.~G. {Andrews}, and F.~K. {Jondral},
  ``Dual-branch {MRC} receivers under spatial interference correlation and
  {Nakagami} fading,'' \emph{{IEEE Trans. on Commun.}}, vol.~62, no.~6, pp.
  1830 -- 1844, Jun. 2014.

\bibitem{arindam:comml:2017}
A.~Ghosh, ``Mixture-based modeling of spatially correlated interference in a
  {Poisson} field of interferers,'' \emph{IEEE Commun. Letters}, vol.~21,
  no.~11, pp. 2496--2499, Nov 2017.

\bibitem{Baccelli:Blaszczyszyn:Muhlethaler2006}
F.~Baccelli, B.~Blaszczyszyn, and P.~Muhlethaler, ``An {Aloha} protocol for
  multihop mobile wireless networks,'' \emph{IEEE Trans. on Inform. Theory},
  vol.~52, no.~2, pp. 421--436, Feb 2006.

\bibitem{mhaenggi:ganti:2009}
M.~Haenggi and R.~K. Ganti, ``Interference in large wireless networks,''
  \emph{Found. and Trends in Netw.}, vol.~3, no.~2, pp. 127--248, 2009.

\bibitem{jhonson:bell}
W.~P. Johnson, ``The curious history of {Fa\'{a}} di {Bruno's} formula,''
  \emph{The American Mathematical Monthly}, vol. 109, no.~3, pp. 217--234,
  2002.

\bibitem{bruno:1857}
F.~di~Bruno, ``Note sur un nouvelle formule de calcul differentiel,''
  \emph{Quarterly J. Pure Applied Mathematics}, vol.~1, p. 359–360, 1857.

\bibitem{haenggi:2012}
M.~Haenggi, ``Diversity loss due to interference correlation,'' \emph{IEEE
  Commun. Letters}, vol.~16, no.~10, pp. 1600--1603, October 2012.

\bibitem{jacobson:2009}
N.~Jacobson, \emph{Basic Algebra-1}, 2nd~ed.\hskip 1em plus 0.5em minus
  0.4em\relax Dover, 2009.

\end{thebibliography}
\end{document}